\documentclass[journal]{IEEEtran}

\DeclareMathSizes{10}{9}{7}{6}   % For size 10 text

\usepackage{setspace}
\usepackage{graphicx}  % Written by David Carlisle and Sebastian Rahtz

\usepackage{amsmath}   % From the American Mathematical Society
\begin{document}

%\textwidth 6.5in
%\textheight 8.6in
%\singlespacing

\title{Underwater Acoustic Networks: Channel Models and Network Coding based Lower Bound to Transmission Power for Multicast }

%HHHHHHHHHHHHHHHEEEEEEEEEEEEEEEERRRRRRRRRRRRRRRREEEEEEEEEEEEEEEEEEEEE

\author{ Daniel E.~Lucani,~\IEEEmembership{Student Member,~IEEE,}
        Muriel~M\'edard,~\IEEEmembership{Fellow,~IEEE,}
       and~Milica~Stojanovic,~\IEEEmembership{Member,~IEEE}
\thanks{Manuscript received March 8, 2008; revised July 15, 2008.
This work was supported in part by the National Science Foundation under grants No. 0520075, 0427502, and CNS-0627021, by ONR MURI Grant No. N00014-07-1-0738, and by United States Department of the Navy's Space and Naval Warfare Systems Command (SPAWAR) under Contract No. N66001-06-C-2020 through BAE Systems.}% <-this % stops a space
\thanks{Daniel E. Lucani and Muriel M\'edard are with the Research Laboratory of Electronics, Massachusetts Institute of Technology, Cambridge, MA 02139 (email: \{dlucani, medard\} @mit.edu).}
\thanks{Milica Stojanovic is with the ECE Department Northeastern University, Boston, MA 02118 (email: millitsa@mit.edu)}}% <-this % stops a space

% note the % following the last \IEEEmembership and also the first \thanks - 
% instance, "A\textbf{} \textbf{}B" will typeset as "A B" not "AB". If you want
% "AB" then you have to do: "A\textbf{}\textbf{}B"
\markboth{Journal of Selected Areas in Communications,~Vol.~1, No.~11,~November~2008}{Lucani \MakeLowercase{\textit{et al.}}: Underwater Acoustic Networks: Channel Models and Network Coding based Lower Bound to Transmission Power for Multicast}
% *** Note that you probably will NOT want to include the author's name in ***
% *** the headers of peer review papers.                                   ***

% If you want to put a publisher's ID mark on the page
% (can leave text blank if you just want to see how the
% text height on the first page will be reduced by IEEE)
%\pubid{0000--0000/00\$00.00~\copyright~2002 IEEE}

%\specialpapernotice{(Invited Paper)}

\maketitle

\IEEEpeerreviewmaketitle

\begin{abstract}

The goal of this paper is two-fold. First, to establish a tractable model for the underwater acoustic channel useful for network optimization in terms of convexity. Second, to propose a network coding based lower bound for transmission power in underwater acoustic networks, and compare this bound to the performance of several network layer schemes. 
The underwater acoustic channel is characterized by a path loss that depends strongly on transmission distance and signal
frequency. The exact relationship among power, transmission band, distance and  capacity for the Gaussian noise scenario is a complicated one.
We provide a closed-form approximate model for 1) transmission power and 2) optimal frequency band to use, as functions of distance and capacity. The model is obtained through numerical evaluation of analytical results that take into account physical models of acoustic propagation loss and ambient noise. 
Network coding is applied to determine a lower bound to transmission power for a multicast scenario, for a variety of multicast data rates and transmission distances of interest for practical systems, exploiting physical properties of the underwater acoustic channel. The results quantify the performance gap in transmission power between a variety of routing and network coding schemes and the network coding based lower bound. We illustrate results numerically for different network scenarios.

\end{abstract}


% Will be ignored for other modes.
\IEEEpeerreviewmaketitle

%\doublespacing
\section{Introduction}
% \PARstart{A}{demo} file is ....
% \PARstart{A}{}demo file is ....
% \PARstart{T}{his demo} file is ....

With recent advances in acoustic communication technology, the interest in study and experimental deployment of underwater networks has been growing \cite{partan06}. However, underwater acoustic channels impose many constraints that affect the design of wireless networks. They are characterized by a path loss that depends on both transmission distance and signal frequency, in a far more pronounced way than a terrestrial radio system. Thus, not only the transmission power, but also the useful bandwidth depend strongly on transmission distance \cite{milica06}. References \cite{sozer00} and \cite{akyildiz05} present studies of the characteristics and design challenges of underwater acoustic networks.  

	In terms of the network layer, a series of routing schemes have been proposed for underwater networks over the recent years. References \cite{sozer00}, \cite{akyildiz05} and \cite{akyildiz06} present surveys of different routing schemes used in underwater networks. In \cite{pompili06} two distributed routing algorithms are introduced for delay-insensitive and delay-sensitive applications. Reference \cite{carlson06} presents a modification of the dynamic source routing protocol that adds location awareness and link quality metrics. In \cite{cui08} a routing protocol based on local depth of the nodes is studied.    

	Network coding was introduced by Ahlswede \textit{et al} \cite{ahslwede00}. Network coding considers the nodes to have a set of functions that operate upon received or generated data packets \cite{lun06}. Work in \cite{li03} and \cite{medard03} showed that linear codes over a network are sufficient to implement any feasible multicast connection. Also, \cite{medard03} provides an algebraic framework for studying this subset of coded networks. Work in \cite{ho06} presents the idea of using linear codes generated randomly in a network. Some practical network coding protocols have been presented in \cite{gerla06} and \cite{chachulski07}. Network coding has previously been considered for underwater networks, showing better performance than other routing schemes. In \cite{guo06} the problem of error recovery was studied in terms of the fraction of delivered packets and total number of transmissions. In \cite{lucani07} routing and network coding schemes were compared based on the time to complete the transmission of a fixed number of packets, and the power required to do so. Existing results compare different network schemes, but the question remains open as to what is the gap between these schemes and the theoretical optimum in terms of transmission power.

	The objectives of this paper are 1) to establish a tractable model for the underwater acoustic channel that will be useful for network optimization in terms of convexity, and 2) to propose a network coding based lower bound for transmission power in underwater acoustic networks by using those models.  The bound is used to compare the performance of several network coding and routing schemes. 
	
	For an underwater acoustic channel both distance between two nodes and capacity determine transmission power and optimal transmission band. However, the complete model that relates these variables is complicated. This paper presents a simple closed-form approximation for transmission power and optimal operating frequency band as functions of distance and capacity. This approximate model stems from an information theoretic analysis that takes into account the physics of acoustic propagation, and colored Gaussian ambient noise. 

	Reference \cite{milica06} shows that transmission power as function of distance can be well approximated by $P(l) = pl^{\gamma}$. A similar model holds for the bandwidth. The coefficients in this model are functions of the required signal to noise ratio (SNR). The present work extends this idea of modeling the power and bandwidth as functions of distance, but the problem is cast into a slightly different framework. Namely, instead of using the SNR as a fixed design constraint, link capacity is used as the figure of merit. The parameters of the approximate model proposed are functions of capacity and distance. This approximate model is useful for a broad range of capacities and distances.

	The complete model that relates transmission power, transmission band, distance, and link capacity is provably convex. Since the approximate model is used instead of the complete model in network optimization, the present work shows the operating conditions under which the approximate model is convex.  	
	
	We assess the minimum transmission power required for an underwater acoustic network. A lower bound for transmission power is obtained by neglecting interference between the nodes and using subgraph selection \cite{lun06} to establish minimum-cost multicast connections with network coding. The convex cost function for the network optimization is given by the transmission power which depends on the distance and a desired data rate via the approximate model for each active link. We show that the no-interference assumption in an underwater scenario is justified for low multicast rates, and randomly placed nodes with inter-node distances less than 10~km. 
		
	Finally, the network coding based lower bound for transmission power is used to compare different routing and network coding schemes. We use some of the schemes in \cite{lucani07} for a concatenated relay network and extend them to a random deployment of nodes in two dimensions. Also, we use an ALOHA-like MAC layer instead of a TDMA scheme as in \cite{lucani07} since TDMA is not scalable. Furthermore, the problem of scheduling in TDMA is NP-complete \cite{ergen05}.

	The paper is organized as follows. In Section \ref{ChannelModel}, the complete model of an underwater channel with Gaussian noise is presented, and special characteristics of this channel are highlighted. In Section \ref{ApproximateModel}, the approximate model for the underwater channel is presented. In Section \ref{ConvexityModel}, convexity of the complete model is proven and conditions for convexity of the approximate model are studied. In Section \ref{OptimizationProblem.tag}, the lower bound for transmission power using network coding and subgraph selection is computed. Section \ref{PerformanceComparison} presents the schemes to be used for performance evaluation and gives numerical results a two-dimensional network scenario. The gap of several routing and network coding schemes to the network coding based lower bound is determined. The last section summarizes the conclusions of this work and future research topics. 
\section{Channel Model} \label{ChannelModel}
An underwater acoustic channel is characterized by a path loss that depends on both distance \textit{l} in km and signal frequency \textit{f} as

\begin{equation}
A(l,f) = (l/l_{ref})^{k}{a(f)}^{l}
\end{equation}
where \textit{k} is the spreading factor, $l_{ref}$ is a reference distance, and $a(f)$ is the absorption coefficient (Figure 1 in \cite{milica06}). The spreading factor describes the geometry of propagation, e.g. $k=2$ corresponds to spherical spreading, $k=1$ to cylindrical spreading, and $k=1.5$ to practical spreading.  The absorption coefficient can be expressed in dB/km using Thorp's empirical formula for $f$ in kHz \cite{berkhovskikh82}, which is an strictly increasing function.

The noise in an acoustic channel can be modeled through four basic sources: turbulence $N_{t}(f)$, shipping $N_{s}(f)$, waves $N_{w}(f)$, and thermal noise $N_{th}(f)$ \cite{milica06}. The power spectral densities (psd) of these noise components in dB re $\mu$ Pa per Hz as functions of frequency in kHz are presented in \cite{coates89}. These psd's have two important parameters: 1) the shipping activity $s$ ranging from 0 to 1, for low and high activity, respectively, 2) the wind speed $w$ measured in m/s. Figure 2 in \cite{milica06} shows $N(f)$ for different values of $s$ and $w$, and an approximation $10 \log N(f) = N_{1} - \eta \log(f) $ for $f \leq 100$~kHz, where $N_{1} = 50$~dB~re~$\mu$~Pa and $\eta = 18$~dB/dec. 

	Assuming that the channel is Gaussian, its capacity can be obtained using the waterfilling principle \cite{cover}\cite{gallager}. The capacity of a point-to-point link is
\begin{align} \label {Capacity_func}
C = \int _{B(l,C)} log_{2} \left (   \frac{K(l,C)}{A(l,f)N(f)}  \right ) df
\end{align} 
where $B(l,C)$ is the optimum band of operation and $K(l,C)$ is a constant determined to satisfy a given constraint.
The band $B(l,C)$ could be thought of as a union of non-overlapping intervals,  $B(l,C) = \cup _i [f_{ini}^i(l,C),f_{end}^i(l,C)]$, where each non-overlapping interval $i$ has the lower-end frequency $f_{ini}^i(l,C)$ and the higher-end frequency $f_{end}^i(l,C)$ associated with it. The factor $1/{A(l,f)N(f)}$ is shown in Figure 3 in \cite{milica06} for different values of $l$. This figure suggests that the optimal transmission band changes considerably with respect to the distance \cite{milica06}. 
The transmission power associated with a particular choice of $(l,C)$ is given by
\begin{align} \label{Power_func}
P(l,C) = \int _{B(l,C)} S(l,C,f) df
\end{align} 
where $ S(l,C,f) = K(l,C) - A(l,f)N(f), f \in B(l,C)$ is the psd of the signal.
	The corresponding signal-to-noise ratio is given by: 
\begin{eqnarray} \label{SNRequation.tag}
SNR = \frac{\int _{B(l,C)} S(l,C,f)A^{-1}(l,f)\, \mathrm{d}f}{\int _{B(l,C)} N(f)\, \mathrm{d}f}
\end{eqnarray} 
We observe that a choice of $(l,C)$ determines implicitly the SNR level. Hence, there is a one-to-one correspondence between the pair $(l,C)$ and the pair $(l,SNR)$. The latter parameterization was used in \cite{milica06} to compute the transmission power and bandwidth representation to ensure a preset SNR, which determines the value of $C$ implicitly. The present analysis focuses on using the former parameterization, i.e. on determining the power and transmission band that ensure a pre-set link capacity. 
\subsection{Dependence on the spreading factor}
	The dependence on the spreading factor $k$ is quite simple. Let us assume that a model for $P(l,C)$ has been developed for a particular value of $k$, i.e. $P(l,C, k)$. To determine $P(l,C, k')$  for $k' \neq k$, note that a change in $k$, the product $A(l,f)N(f)=(l /l_{ref})^ka(f)^lN(f)$ constitutes a constant scaling factor with respect to $f$. Therefore, for a link of distance $l$ the term $B(l,C)$ will remain unchanged. Thus, if the same capacity $C$ is required for $k$ and $k'$, equation \eqref{Capacity_func} shows that the only other term that can vary is $K(l,C)$, i.e. $K(l,C,k)$. Then, $K(l,C,k') = (l /l_{ref})^{k' - k} K(l,C,k)$. Finally, let us use the equation \eqref{Power_func} to determine the relationship between $P(l,C,k)$ and $P(l,C,k')$. 
\begin{align}
&P(l,C,k') = \int _{B(l,C)} \left( K(l,C,k') - l_{m}^{k'}a(f)^lN(f) \right ) df\\ 
&= l_{m}^{k' - k} \int _{B(l,C)} \left(  K(l,C,k) - l_{m}^{k} a(f)^lN(f) \right ) df\\ 
&= l_{m}^{k' - k} P(l,C,k) \label{SpreadFactorRelation}
\end{align} 	
where $l_{m} = l /l_{ref}$ to shorten the derivation.
	Thus, any model generated for some parameter $k$ has a simple extension. Also, note that the transmission band remains the same for any value of $k$.

\begin{figure}[t]
\centering						
\includegraphics[height=3.5in,width=3.5in]{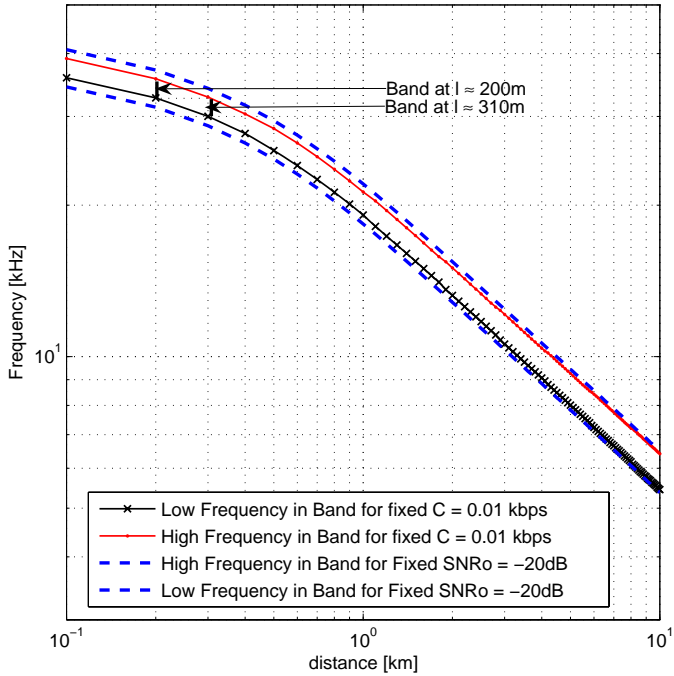}
\caption{High and low band edge frequency of the transmission band for $C = 0.01$kbps, $k=1.5$, $s = 0.5$, and $w = 0 $m/s.   }
\label{HighLowFreqBandC0_01kbps.tag}
\end{figure}    	

\begin{figure}[t]
\centering						
\includegraphics[height=3.5in,width=3.5in]{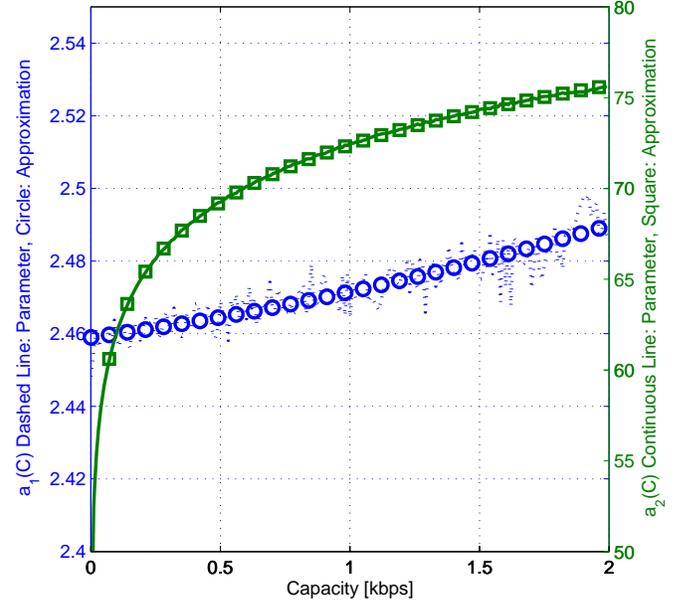}
\caption{Parameters $a_{1}$ and $a_{2}$ for $P(l,C)$ and approximate model.$l \in [0, 10] $~km, $C \in [0,2] $~kbps,$k=1.5$,$s = 0.5$ and $w = 0 $~m/s. }
\label{figPLC_k_15_lowC.tag}
\end{figure}    	

\subsection{Interference Characteristics}
	The optimal transmission band of a link was shown to change with the distance, under the assumption that the channel is Gaussian and that the capacity of a link is obtained through waterfilling. If the capacity for a link is low, e.g. less than 2~kbps, and the transmission distance is below 10~km, the transmission bandwidth will also be low, and its optimal location in the spectrum will change dramatically with the distance. Figure \ref{HighLowFreqBandC0_01kbps.tag} shows this effect for $C = 0.01$~kbps. In this figure, the high and low band edge frequencies are plotted. This figure also shows the high and low band edge frequency if an SNR requirement of -20~dB is set, i.e. using the SNR instead of the capacity as the fixed parameter. As noted before, the constraint over the capacity is related to different SNR levels depending upon the distance. It is clear that low values of $C$ are related to a very low SNR value.

	Figure \ref{HighLowFreqBandC0_01kbps.tag} shows that if two links with the same $C = 0.01$~kbps are established, one with $l \approx 200$~m and the other with $l \approx 310$~m, the optimal transmission bands for these links will not overlap; thus, they do not interfere with one another. This characteristic of the underwater channel suggests that if a network is established in which the nodes are at different distances from one another, and each node has a limited range of transmission when the data rate requirement is very low (all valid assumptions in underwater networks),  there will be no interference between transmissions of the various links. If each link allocates its band optimally, this suggests that a form of FDMA is the optimal approach in an underwater network, where transmission band is determined by both the distance and the required data rate. 
	From a network optimization view point, the cost function to be minimized is clearly separable under these assumptions, where the channel model for a link can be used as the cost function for each of the separable terms.
\subsection{Numerical Evaluation Procedure}
	A numerical evaluation procedure similar to that of \cite{milica06} is used to compute the value of $P(l,C)$, $\hat{B}(l,C)$ and $\hat{f}_{end}(l,C)$, for a region of values of $(l,C)$. The procedure starts by fixing a target value of the capacity $C$. Then, for each distance $l$, the initial value of $K(l,C)$ is set to the minimum value of the product $A(l,f)N(f)$, i.e. $K(l,C) = \min_{f} A(l,f)N(f)$. The frequency at which this occurs, i.e. $f_{0} = \arg \min_{f} A(l,f)N(f)$, is called the optimal frequency. After this, $K(l,C)$ is increased iteratively by a small amount, until the target capacity value $C$ is met. Finally, this procedure is repeated for each value of $C$ in a range of interest.
	At the $n$-th step of the procedure, when $K^{(n)}(l,C)$ is increased by a small amount, the band $B^{(n)}(l,C)$ is determined for that step. This band is defined as the range of frequencies for which the condition $A(l,f)N(f) \leq K^{(n)}(l,C)$ holds. Then, the capacity $C^{(n)}$ is numerically determined for the current $K^{(n)}(l,C)$ and $B^{(n)}(l,C)$, using the equation \eqref{Capacity_func}. If $C^{(n)} < C$, a new iteration is performed. Otherwise, the procedure stops.

\begin{figure}[t]
\centering						
\includegraphics[height=3.5in,width=3.5in]{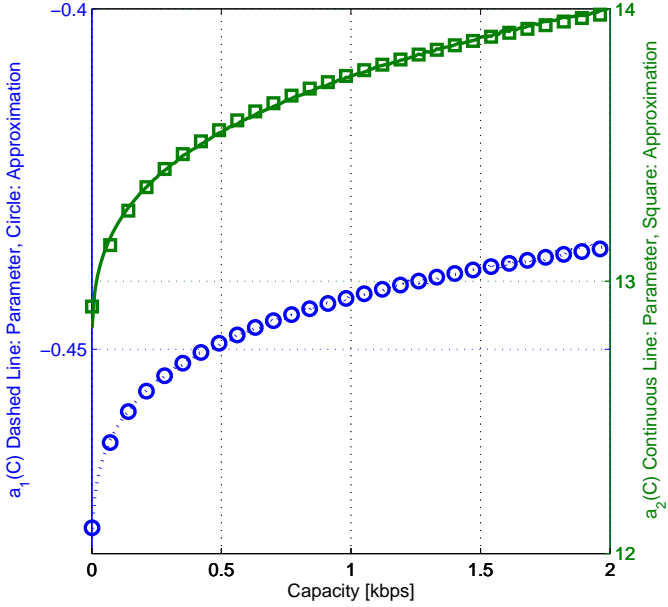}
\caption{Parameters $a_{1}$ and $a_{2}$ for $\hat{f}_{end}(l,C)$ and approximate model.$l \in [0, 10] $~km, $C \in [0,2]$~kbps,$k=1.5$,$s = 0.5$ and $w = 0 $~m/s.  }
\label{FhighLC_k_15_lowC.tag}
\end{figure}
\begin{figure}[t]
\centering	
\includegraphics[height=3.5in,width=3.5in]{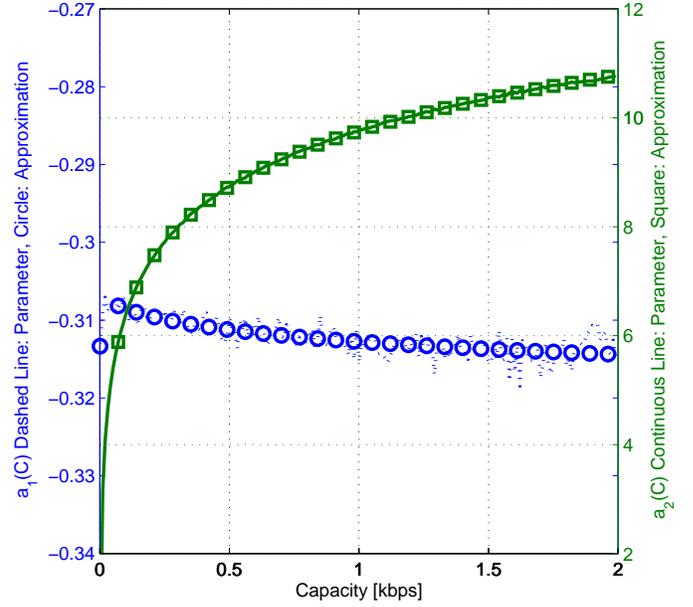}
\caption{Parameters $a_{1}$ and $a_{2}$ for $\hat{B}(l,C)$ and approximate model.$l \in [0, 10] $~km, $C \in [0,2]$~kbps, $k=1.5$,$s = 0.5$ and $w = 0 $~m/s.  }
\label{BLC_k_15_lowC.tag}
\end{figure}

\section{Approximate model} \label{ApproximateModel}
	Evidently, the expressions for the complete model are quite complicated to be used in a computational network analysis. Also, they provide little insight into the relationship between power consumption, $\hat{B}$ and $\hat{f}_{end}$ , in terms of the pair $(l,C)$. This motivates the need for an approximate model to represent these relations for ranges of $C$ and $l$ that are of interest to acoustic communication systems. The model should also provide flexibility to changing other parameters, such as the spreading factor $k$, wind speed $w$ and shipping activity $s$. As shown in Equation \eqref{SpreadFactorRelation}, any approximate model for the transmission power generated for some parameter $k$ has a simple extension to any other value of $k$. Also, a model for the transmission band remains the same for any value of $k$.
	
	By applying the numerical procedure of the previous section for various $l$ and $C$ and fitting the data, it is possible to obtain approximate models for power consumption (Eq.~\ref{PowerApprox}), band-edge frequency $\hat{f}_{end}(l,C)$ (Eq.~\ref{HighFreqApprox}), and for the bandwidth $\hat{B}(l,C) = \hat{f}_{end}(l,C) - \hat{f}_{ini}(l,C)$ (Eq.~\ref{BandApprox}). Note that some important properties for these parameters are kept, e.g. $P(l,0) = 0$. 
	
\singlespacing
\begin{align} \label{PowerApprox}
&\tilde {P}(l,C) = l^{a_{1}(C)} 10^{\frac{a_{2}(C)}{10}} \text{ with } \\
&a_{1}(C) = \alpha _3 + \alpha _2 C  + \alpha _1 C^2\\
&a_{2}(C) =  \beta _3 + \beta _2 10log_{10}C  + \beta _1 (10log_{10}(C+1))^2 \notag
\end{align} 
\begin{align} \label{HighFreqApprox}
&\hat{f}_{end}(l,C) = l^{a_{1}(C)}10^{\frac{a_{2}(C)}{10}} \text{ with }\\
&a_{1}(C) = \alpha _3 + \alpha _2 10log_{10}C  + \alpha _1 (10log_{10}C)^2\\
&a_{2}(C) =  \beta _3 + \beta _2 10log_{10}C  + \beta _1 (10log_{10}C)^2 \notag
\end{align}
\begin{align}\label{BandApprox}
&\hat{B}(l,C) = l^{a_{1}(C)}10^{\frac{a_{2}(C)}{10}} \text{ with }\\
&a_{1}(C) = \alpha _4 + \alpha _3 10log_{10}C + \alpha _2 (10log_{10}C)^2+\alpha _1 (10log_{10}C)^3\\
&a_{2}(C) =  \beta _3 + \beta _2 10log_{10}C  + \beta _1 (10log_{10}C)^2 \notag
\end{align}
%\doublespacing
The transmission power, band-edge frequency and bandwidth of transmission band were computed for a variety of values of $s$, $w$ and two ranges of interest of the pair $(l,C)$: $l \in (0, 10] $~km, $C \in [0,2]$~kbps, and $l \in (0, 100]$~km, $C \in [0,100] $~kbps. The models proposed fitted these cases quite well. Results are presented for $k = 1.5$, $w = 0$ and $s = 0.5$, for both cases. For the first case, the $\alpha$ and $\beta$ parameters show almost no dependence on the shipping activity factor $s$, especially if the wind speed is $w>0$. Thus, the approximate model for this case can be simplified to  consider $w$ only as part of the model, instead of the pair $(s,w)$.

	Figures ~\ref{figPLC_k_15_lowC.tag}, ~\ref{FhighLC_k_15_lowC.tag}  and ~\ref{BLC_k_15_lowC.tag} show parameters $a_{1}$ and $a_{2}$ for $P(l,C)$, $\hat{f}_{end}(l,C)$, and $\hat{B}(l,C)$, respectively, for the first case with $k=1.5$, $s = 0.5$ and $w = 0 $~m/s. The values of $\alpha$'s and $\beta$'s are shown in Table ~\ref{Table_a2_lowC} as Case 1, for parameters $a_{1}$ and $a_{2}$, respectively. These tables also show the mean square error (MSE) of the approximation with respect to the actual parameters. 
	A similar result can be found for the second case with $k=1.5$, $s = 0.5$ and $w = 0$. The values of $\alpha$'s and $\beta$'s are shown as Case 2 in Table ~\ref{Table_a2_lowC}, for parameters $a_{1}$ and $a_{2}$, respectively. For both ranges, the proposed models give a very good approximation to the actual numerical values. Also note that for the $a_1(C)$ parameter of $P(l,C)$, it is possible to use a linear approximation, instead of a quadratic model.
	
	For case 1, the values for $\alpha$ and $\beta$ parameters in the approximate $\tilde {P}(l,C)$ model show very little dependence with respect to $s$ while they show a greater dependency on $w$. This is not unexpected, since the transmission band is at a high frequency (between 5 and 40 KHz) where the noise psd is influenced more by the $w$ ($O(f^{-2})$) than $s$ ($O(f^{-3.4})$). Therefore, a further approximation is to discard $s$ and consider parameters $\alpha$ and $\beta$ to be functions of $w$ only. In particular, a simple model is $ \alpha_i = \gamma _3 + \gamma _2 10log_{10}(w+1)  + \gamma _1 (10log_{10}(w+1))^2$, $\forall i = 1,2,3$. A similar relation holds for $\beta_i, \forall i = 1,2,3$. Table ~\ref{Table_windspeedrelation_lowC} shows $\gamma$ parameters for the different $\alpha$'s and $\beta$'s.

\begin{table}[10pt,t]
\onecolumn
\caption{$a_{1}$ and $a_{2}$ approximation parameter values for $P(l,C)$, $\hat{f}_{end}(l,C)$ and $\hat{B}(l,C)$, with  $k=1.5$,$s = 0.5$ and $w = 0 $~m/s for Case 1: $l \in [0, 10]$~km, $C \in [0,2]$~kbps, and Case 2: $l \in [0, 100] $~km, $C \in [0,100] $~kbps.}
\centering
\label{Table_a2_lowC}
\scriptsize
\begin{tabular}{|c|c||c|c|c|c|c||c|c|c|c|}
\hline
 Case&&$\alpha _1$&  $\alpha _2$    &   $\alpha _3$  & $\alpha _4$&MSE &  $\beta _1$    &   $\beta _2$  & $\beta _3$ & MSE\\
\hline
1 &$P(l,C)$    & -0.00235 & 0.01565 & 2.1329 & 0 &2.532e-7  & 0.014798 & 1.0148 & 74.175 & 5.8979e-5\\
\hline
1&$\hat{f}_{end}(l,C)$   & 4.795e-5 &0.00246&-0.44149 & 0 & 3.930e-9  & 0.00171 & 0.07153 & 13.738 &3.4706e-5  \\
\hline
1&$\hat{B}(l,C)$  &-5.958e-7 & -2.563e-5 & -0.000305 & -0.30694  & 6.599e-9 &  -5.163e-6 & 0.33427 & 9.6752 & 2.9233e-7  \\
\hline
\hline
2&$P(l,C)$   & -5.617e-5 & 0.02855 & 2.9305& 0& 0.00011   & 0.04317 & 0.90597 & 76.156 & 0.00010115\\
\hline
2&$\hat{f}_{end}(l,C)$   & -0.00019 &0.01186 &-0.55076 & 0& 1.32e-7& 0.0065157 & -0.032693 & 14.739 &7.3024e-5 \\
\hline
2&$\hat{B}(l,C)$  &1.696e-6 & 4.252e-5 & -0.00249 & -0.36397 &7.29e-7 & -0.0018252 & 0.34788 & 10.328 & 0.00019414\\
\hline
\end{tabular}				
\normalsize
\twocolumn
\end{table}

\section{Convexity Analysis} \label{ConvexityModel}
 In this section we prove the convexity of the complete model of $P(l,C)$ in the entire region of interest, i.e. positive data rates and $l>0$. Then we discuss the necessary conditions that the value of $l$ has to fullfill to ensure convexity of the approximate model.
\subsection{Convexity of Complete Model}
	The convexity of the transmission power of the complete model is stated in the following lemma, which is proven in the Appendix. Lemma 1 assures that $P(l,C)$ is a convex function with respect to $C$ for the ranges of interest of $C$ and $l$ for the case of non-overlapping finite bands.

	\textit{Lemma 1} $P(l,C)$ is a convex, increasing function with respect to $C$, $\forall C>0$ and $l>0$, if $A(l,f)N(f) > 0$ and $B(l,C) = \cup _i [f_{ini}^i(l,C),f_{end}^i(l,C)]$, with $f_{ini}^i(l,C)<f_{end}^i(l,C)< \infty, \forall i $ and  $f_{ini}^i(l,C), f_{end}^i(l,C) \not \in [f_{ini}^j(l,C),f_{end}^j(l,C)], \forall i \neq j, $ i.e. a union of non-overlapping finite bands.
	\textit{Proof} See Appendix I.

\begin{tabular}[h]{c}
                 \\
                \\
                \\
                 \\
                \\
                \\
                 \\
                \\
                \\
                 \\
                \\
                \\                
\end{tabular}

	Another property to be used is given in Lemma 2. This lemma assures that if a link between a transmitter $i$ and receiver $j$ at a distance $l$ achieves a certain capacity $C$, another node $k$ at distance $l' < l$ from node $i$, will be able to decode the information transmitted from $i$ to $j$. Note that the transmission band is optimal for the link of distance $l$.

	\textit{Lemma 2} $C(l,B(l,C)) <  C(l',B(l,C))$ for $l' < l$, if $A(l,f) = (l /l_{ref})^k{a(f)}^l$ with $k \geq 1$ and $a(f) \geq 1, \forall f$.
	\textit{Proof} See Appendix II.

\subsection{Convexity of Approximate Model}
	The function $P(l,z)$ represents the minimum power required to transmit at a data rate $z$ over a link of distance $l$. The function $P(l,z)$ was proven to be a convex function with respect to $z$, using $l$ as a parameter (Lemma 1). However, the exact model is complicated from a computational viewpoint. Let us determine the conditions for which the approximate model $\tilde{P}(l,z)$ in equation \eqref{PowerApprox} is convex with respect to $z$, and having $l$ as a fixed parameter. We study the case of $z<2$ kbps. 
	Since the $\alpha$ and $\beta$ parameters come from fitting the data, the only variable left to analyze is the distance $l$. Note that ensuring that $\tilde{P}(l,z)$ is increasing and convex translates into the following inequalities:
%\singlespacing
\begin{align}
&ln(l) \scriptstyle \frac{\partial a_{1}(z)}{\partial z} \displaystyle + \scriptstyle \frac{ln(10)}{10}  \frac{\partial a_{2}(z)}{\partial z} \displaystyle >0 \label{LinConstraint}\\
&ln(l)^2 \left ( \scriptstyle \frac{\partial a_{1}(z)}{\partial z } \displaystyle \right )^2 +
 \scriptstyle \frac{ln(10)}{10} \frac{\partial ^2 a_{2}(z)}{\partial z^2 } \displaystyle + \left( \scriptstyle \frac{ln(10)}{10}  \frac{\partial a_{2}(z)}{\partial z } \displaystyle \right )^2 \notag \\ 
&+ ln(l) \left( 2 \scriptstyle \frac{ln(10)}{10} \frac{\partial a_{1}(z)}{\partial z }\frac{\partial a_{2}(z)}{\partial z } \displaystyle +  \scriptstyle \frac{\partial ^2 a_{1}(z)}{\partial z^2 } \displaystyle \right)  \geq 0 \label{QuadConstraint}
\end{align}
%\doublespacing
There is both a linear and a quadratic constraint upon $l$ to ensure convexity. Since these constraints are also functions of $z$, the range of values of this parameter should be considered. From previous results for the fitting parameters, it is possible to determine some properties of the model for $z < 2 $ kbps. In terms of the parameters of interest, $\alpha _1<0$, $\alpha _2 >0$, $2 \alpha_1 C + \alpha_2 >0$, $\beta_1 >0$ and $\beta_2 >0$.  Thus, for the choices of $a_{1}(z)$ and $a_{2}(z)$, the first and second derivatives of these functions with respect to $z$ are $\dot {a_{1}}(z) > 0$, $\ddot {a_{1}}(z) < 0$, $ \dot {a_{2}}(z)> 0$ $\ddot {a_{2}}(z) < 0$. Using these conditions, the constraints \eqref{LinConstraint} and \eqref{QuadConstraint} can be simplified to
\begin{align} 
&ln(l) > - \scriptstyle \frac{ln(10)}{10} \displaystyle \frac{\dot {a_{2}}(z)}{\dot {a_{1}}(z)} + 
\max [ 0, - \frac{\ddot {a_{1}}(z)}{2 {\dot {a_{1}}(z)}^2} + \\&
\sqrt { \scriptstyle \frac{ln(10)}{10} \displaystyle \left ( \frac{\dot {a_{2}}(z) \ddot {a_{1}}(z)}{ {\dot {a_{1}}(z)}^3} -   \frac{ {\ddot {a_{2}}(z)}}{{\dot {a_{1}}(z)}^2} \right) + \frac{{\ddot {a_{1}}(z)}^2}{4 {\dot {a_{1}}(z)}^4 }   } ]
\end{align} 		 
where the term under the square root is positive which ensures real values of $l$. Note that for each value of $z$ there is a minimum value of $l$. Let us use the values of Case 1 in Table ~\ref{Table_a2_lowC} to determine the $(l,z)$ region for which the approximate model is convex. For these values, if the distance between to nodes $l$ is at least 13~m, for any value of $z <$2~kbps the model will be convex. The limitation to $l>$13 m is related to the sampling of the distance used for computing the parameters of the approximate model. For all practical purposes the approximate model $\tilde{P}(l,z)$ is convex.   

\begin{table}[10pt]
\caption{Approximation parameters of $\alpha$ and $\beta$ for $P(l,C)$, $l \in [0, 10] $~km, $C \in [0,2]$~kbps, $k=1.5$,$s = 0.5$}
\centering
\label{Table_windspeedrelation_lowC}
\scriptsize
\begin{tabular}{|c||c|c|c|}
\hline
 &  $\gamma _1$    &   $\gamma _2$  & $\gamma _3$\\
\hline
$\alpha_1$   & 5.2669e-6 & -0.000157 & -0.004575 \\
\hline
$\alpha_2$   & -2.971e-5 & 0.000865 & 0.029306 \\
\hline
$\alpha_3$   & 0.000152 & 0.01809 & 2.4586\\
\hline
$\beta_1$   & 9.924e-6 & -0.00027 & 0.012288 \\
\hline
$\beta_2$   & 7.799e-6 & -0.000219 & 1.0118 \\
\hline
$\beta_3$   & 0.068091 & 1.3659 & 73.144\\
\hline
\end{tabular}				
\normalsize
\end{table}

\section{Lower Bound to Transmission Power in Underwater Networks} \label{OptimizationProblem.tag}
	The problem of achieving minimum-energy multicast using network coding in a wireless network has been studied previously \cite{lun06}. A wireless network, as presented in \cite{lun06} can be represented through a directed hypergraph $H = (\aleph , A)$ where $\aleph$ is the set of nodes and $A$ is the set of hyperarcs. A hypergraph is a generalization of a graph, where there are hyperarcs instead of arcs. A hyperarc is a pair $(i,J)$, where $i$, the start node, is an element of $\aleph$ ,and $J$ is the set of end nodes is a nonempty subset of $A$. Each hyperarc $(i,J)$ represents a broadcast link from node $i$ to nodes in the nonempty set $J$. Let us denote by $z_{iJ}$ the rate at which coded packets are injected into hyperarc $(i,J)$. 
If the cost function is separable, the optimization problem can be expressed as follows

\singlespacing
\begin{align}
& \min  \sum_{(i,J)\in A} \theta f(z_{iJ}/ \theta) \notag\\
&\text{subject to }z \in Z \notag\\
&z_{iJ} \geq \sum_{j \in J} x^{(t)}_{iJj}, \forall (i,J) \in A, t \in T\notag\\
&\sum _{\{ J|(i,J)\in A \} } \sum _{j \in J} x^{(t)}_{iJj} - \sum _{ \{ j|(j,I) \in A \}, i \in I} x^{(t)}_{jIi} = \delta ^{(t)} _{i}\notag\\
&x^{(t)}_{iJj} \geq 0, \forall (i,J) \in A, j \in J, t \in T \label{OptimizationNetCod}
\end{align}
with 
\begin{align}
\delta ^{(t)} _{i} =
\begin{cases}
 R & \text{if $i=s$,}\\
-R & \text{if $i=t$,}\\
0 & \text{otherwise}
\end{cases}
\end{align}
%\doublespacing
where $T$ is a non-empty set of sink terminals, a source $s$, a multicast rate $R$, and a fixed transmission duty cycle at each link $\theta$. $x^{(t)}_{iJj}$ represents the flow associated with terminal $t$, sent through hyperarc $(i,J)$ and received by node $j \in J$. 

	In the underwater scenario this formulation is used to establish a lower bound on the transmission power required to achieve a multicast rate $R$. Assuming no interference for transmissions in different hyperarcs yields a separable cost function. Note that if interference was taken into account, the power to reach the desired data rate would increase. Then, the cost function $ f(z_{iJ})$ for each particular hyperarc corresponds to a link transmission power $P(l,z_{iJ})$ in order to obtain the minimum transmission power required to achieve a data rate of $z_{iJ}$, where $l$ represents the distance from $i$ to the farthest node $j \in J$. For the lower bound computation, continuous transmission ($\theta = 1$) is assumed. 
	A simplification of this problem can be made under the assumption that transmissions are omnidirectional, and considering the fact that if a node transmits over a certain range, all nodes in that range will be able to receive the information. This was proven in Lemma 2. Finally, the model for this channel ensures that any value of $z_{iJ}$ can be achieved if enough power is used. Thus, the constraint set $Z$ can be dropped.
	
	Although the problem for minimum-cost multicast is well known for wireless radio networks, the cost function presented here is different because it represents the minimum transmission power for an hyperarc transmitting at
a data rate $Z$, which is given by the power needed to transmit at capacity $C = Z$, without assumption on technology or, more importantly, a specific transmission band which is usually the case for wireless radio networks. Thus, we are providing a lower bound valid for any acoustic underwater network for the case of Gaussian noise.

\section{Performance Comparison} \label{PerformanceComparison}
	For this study, five schemes are considered. The first scheme corresponds to the lower bound to the transmission power using network coding given by solving the problem in Section \ref{OptimizationProblem.tag} with $\theta = 1$. The second scheme corresponds to solving the problem in Section \ref{OptimizationProblem.tag} for $\theta <1$, in order to study the effect of using a duty cycle for link transmissions in underwater networks over interference and transmission power. The third scheme corresponds to using the paths chosen by the optimal scheme but establishing a SNR requirement for the transmission links with the objective of studying interference when the SNR requirement is increased. The links are considered to transmit continuously. The schemes (4) and (5) consider implementations of network coding in a rateless fashion with the implicit acknowledgment (ACK) \cite{lucani07} and routing with link-by-link ACK using an ALOHA-like MAC layer. 
	Let us explain in more detail each of the schemes. 
	
\textbf{1)Network coding based lower bound to transmission power:} Transmission power is computed by solving the convex optimization problem in \ref{OptimizationProblem.tag} and it provides a lower bound on the optimal transmission power for networks operating at low data rates. For this computation, continuous transmission, i.e. a duty cycle of $\theta = 1$ is used. This scheme is used as the gold standard to which the remaining schemes are compared. The no-interference assumption is assessed by computing the average percent of the randomly deployed networks that have at least a link which suffers from severe interference, i.e. a signal-to-interference ratio (SIR) below 3~dB. 

\textbf{2)Network Coding with optimal power consumption for links with fixed duty cycle:} Transmission power is computed by solving the convex optimization problem in \eqref{OptimizationProblem.tag} for links with a fixed duty cycle, i.e. $\theta < 1$. This value provides a lower bound on the optimum power consumption for networks operating at low data rates when links have a particular duty cycle. By convexity of the cost functions used, this bound will be higher than for the previous scheme. This scheme is used to illustrate the effect upon transmission power and interference when the links transmit at a fixed duty cycle $\theta < 1$ by comparing this scheme to the previous scheme. 

\textbf{3)Network Coding with SNR requirement on link transmission:} This scheme is a heuristic scheme that imposes an SNR requirement for transmissions. Using the subgraph selected by solving the problem in \eqref{OptimizationProblem.tag} for $\theta = 1$, it computes the SIR on the different links for a variety of SNR constraints using the models of transmission power and band in \cite{milica06}. As in scheme (1) and (2), the percent of randomly deployed networks with at least a link with severe interference is computed. Results of this scheme suggest that continuous transmission with a moderate SNR requirement causes severe interference. A solution to this problem is to use of a duty cycle $\theta < 1$, similarly as in scheme (2) when there is an SNR requirement.

\begin{figure}[t]
\centering						
\includegraphics[height=2.5in,width=3.5in]{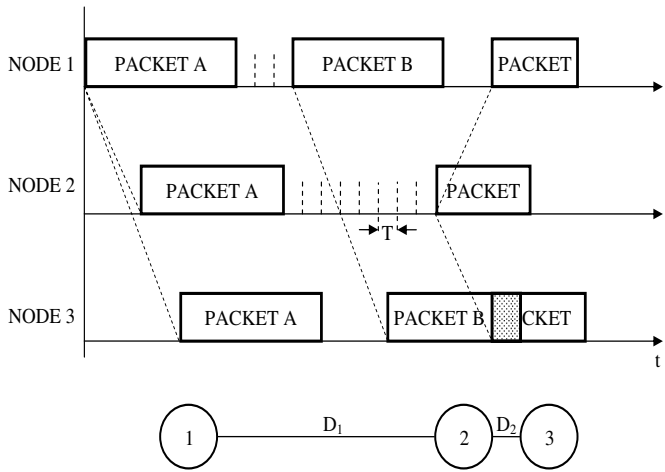}
\caption{Medium Access Protocol for schemes (4) and (5) }
\label{MAC.tag}
\end{figure}

\begin{figure}[t]
\centering						
\includegraphics[height=1in,width=2.5in]{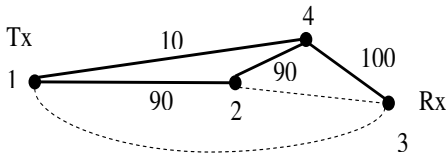}
\caption{Subgraph selection for scheme (4). Selected subgraph with values of $z_{iJ}$ to provide a unicast rate of 100, where dashed lines represent unused transmission ranges. Note that hyperarcs include in the same range, e.g. the possible hyperarcs with node 1 as the staring point are $1\{2\}$,$1\{2,4\}$,$1\{2,3,4\}$. }
\label{Subgraph.tag}
\end{figure}

\textbf{4)Network coding in rateless fashion with implicit ACK:} For a concatenated relay network as in Figure \ref{MAC.tag}, the path between a source node and sink node is fixed and includes all relay nodes. If a node $b$ is closer than node $a$ to the collecting node, $a$ is said to be upstream with respect to $b$, and $b$ is said to be downstream with respect to $a$. For the concatenated relay network this ordering is quite natural. This problem was studied in \cite{lucani07}. 
For a two-dimensional scenario, subgraph selection \cite{lun06} with linear and separable cost functions are used to determine the active links in the network and the transmission power required for each link. The cost function of each hyperarc is computed based on the approximate formulas for transmission power and bandwidth for a fixed SNR level \cite{milica06}. The weight of each link is given by $D P(l,SNR)/B(l,SNR)$, where $D$ is a constant common to all links related to the number of transmitted bits per burst and modulation used. For the performance computation using optimal modulation, i.e. Gaussian signaling, the weight for each link in the path is $ D P(l,SNR)/C(l,SNR)$, where $C(l,SNR)$ is the function of capacity related to the pair $(l,SNR)$. We assume that the coding is over very large number of data packets. This could be extended computing an error probability based on error exponents \cite{gallager}.
Once the subgraph has been selected, if several links share the same transmitting node, this node will randomly choose the link to use. The weight of each link in the random choice is given by the fraction of data rate the optimization problem assigned to each of these links. Let us consider the network in Figure~\ref{Subgraph.tag} as an example, where dashed lines represent unused hyperarcs. If node 1 transmits to both node 2 and 4, but we send a rate of 90 units through hyperarc $z_{1{2}}$, while we send 10 units through hyperarc $z_{1{2,4}}$, then when node 1 transmits it will do so 90~\% of the time to reach node 2 only, and 10~\% of time using enough power to reach nodes 2 and 4. Finally, we have to determine which nodes are upstream and downstream to each node in the subgraph. If the subgraph corresponds to a single path, the choice is clear. If there are multiple paths, we use the following heuristics: we start by ordering the nodes starting at the transmitter and looking at the nodes directly connected to it in the optimal subgraph. These nodes are ordered as follows: the node associated with the link with higher data rate from the transmitter is considered to be directly downstream from the source node, the node with the second highest data rate is considered to be downstream with respect to the previous one, and so on. In Figure~\ref{Subgraph.tag}, 2 is directly downstream of 1, and 4 is downstream of 2. Once all nodes connected to the transmitter (let us call this set of nodes $S$) are ordered, we proceed to order the nodes connected to $S$ by a similar procedure as for the case of one node described before. In the example, $S =\{2,4\}$ and the nodes connected to it are $\{3,4\}$. If a node connected to one of the nodes in $S$ has already been ordered, like node 4 in the example, the link is discarded keeping the previous order of the nodes. We update $S$ with the nodes that were connected to $S$ and not previously in it, until we reach the receiver. For the network example in Figure~\ref{Subgraph.tag} the ordering is 1,2,4,3.

For this particular scheme, once a relay node gets its first coded packet, i.e. a packet formed by a random linear combination of data packets, it will transmit until the receiving node sends a confirmation that all the information has been received. The same happens at the source node. However, nodes eavesdrop on other transmissions. If a node receives a coded packet from a node further downstream with the same, or a greater number of degrees of freedom than what it has, it will stop transmitting and update its information if necessary. Degrees of freedom in this setting represents the number of packets that were linearly combined to form the coded packet as in \cite{lucani07}. The node will resume transmitting if an innovative packet, i.e. a packet with a new random linear combination of data packets useful for decoding the information, is received from a node upstream. The sink node will retransmit a coded packet with its degrees of freedom when a coded packet is received. This strategy assumes that there is a mechanism that informs the collecting node about the number of degrees of freedom that constitute the total message or that this number is fixed \textit{a priori}. 

\textbf{5)Routing using link-by-link acknowledgement:} For a concatenated relay network, the path between the source node and the sink node is fixed and includes all relay nodes. This problem was studied in \cite{lucani07}. For a two dimensional scenario, the sink and the source are chosen randomly and the shortest path is computed before starting data transmission in unicast. The weight of each link is computed based on the approximate formulas for transmission power and bandwidth for a fixed SNR level in the same fashion as the cost function per link of scheme (4). 
In the current scheme, every time a node receives a packet, it will retransmit the packet and send an acknowledgement to the previous node. Once a packet has been acknowledged, the node can start transmitting a new data packet in its queue. If it has no new packets to transmit, it will only transmit if a node upstream sends new information, or sends a previous packet, in which case the node will acknowledge this packet. 

	In terms of the physical layer, schemes (4) and (5) use both PSK modulation, which implies the use of a data rate in each link that is lower than capacity, and Gaussian signaling assuming that the encoding is over a large number of bits. In order to deal with the SNR requirement, we use an approximate model for the transmission power, high band edge frequency and bandwidth as functions of SNR similar to the work in \cite{milica07relay}. When PSK modulation is used, probability of packet error due to noise over the link from node $i$ to $j$ is obtained from the probability of bit error by $P_{\text{packet Error}}(i,j) = 1 - \left( 1 - P_{\text{bit error}}\right)^n$, where $n$ is the number of bits in the packet, and $P_{\text{bit error}}$ is computed using the standard PSK bit error probability. Note that nodes farther away from the transmitter have some probability of receiving the packet correctly. For Gaussian signaling, the probability of packet error due to noise is considered to be zero for all nodes in range, and 1 for all nodes further away.
	
	In terms of the MAC layer, schemes (4) and (5) use an ALOHA-like MAC layer. This ALOHA protocol considers a fixed number of bits per data packet and uses the optimal transmission band for an SNR requirement per link. Thus, the duration of the transmitted packet depends on the transmission distance \cite{milica06}. Every node has a probability to access the medium every $T$ units of time following a Bernoulli process. Transmission delay is considered using a typical value of sound speed (1500~m/s). Figure \ref{MAC.tag} shows an example of using this MAC layer for three nodes with $D_{1} >> D_{2}$. In this example, when node 1 transmit a packet to node 2, this packet also reaches node 3. Note that the duration of the packet transmitted from node 1 to node 2 (Packet A) is large compared to the packet transmitted from node 2 to node 3 because of the relation of distance to bandwidth/capacity for a fixed SNR value mentioned above \cite{milica06}. Once node 1 has transmitted the packet it will try to transmit again, and it has some probability to start transmission every time slot $T$. Let us assume that node 2 has a data packet for node 3. Figure \ref{MAC.tag} shows the case when the data packet transmitted from node 2 to node 3 suffers a collision at node 3 with a new packet transmitted from node 1 to node 2. We consider that a collision at any receiver causes a loss of all packets involved in the collision for that receiver.        
	
\begin{figure}[t]
\centering						
\includegraphics[height=3.5in,width=3.5in]{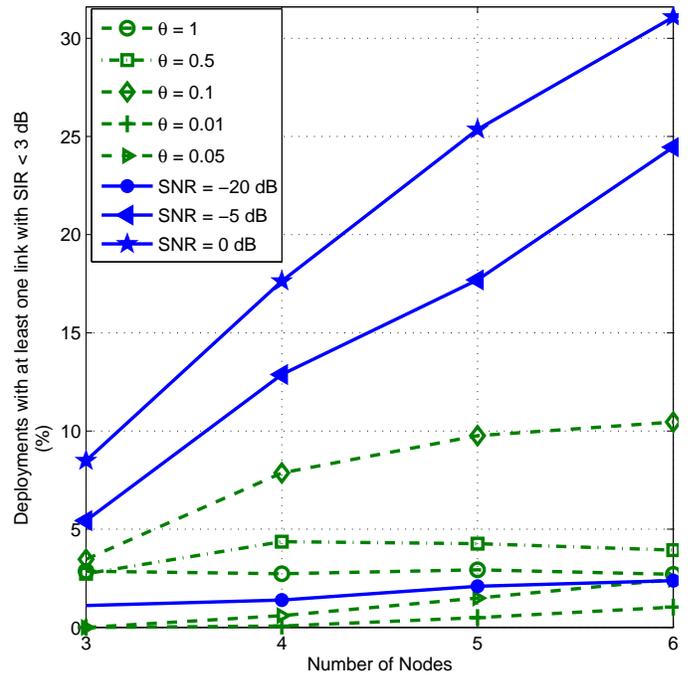}
\caption{Percent of deployments in a fixed square of 5x5 km$^2$ with $SIR < 3$~dB in at least one link vs number of nodes deployed in that area, for the first three schemes. For scheme 1 $\theta= 1$, while scheme 2 is shown with different values of $\theta$ to achieve unicast rate of $R = 0.1$~kbps. Performance for scheme 3 is shown for different SNR values.}
\label{PercentSameSquareDifferentOmega_andSNR.tag}
\end{figure}	

\begin{figure}[t]
\centering						
\includegraphics[height=3.5in,width=3.5in]{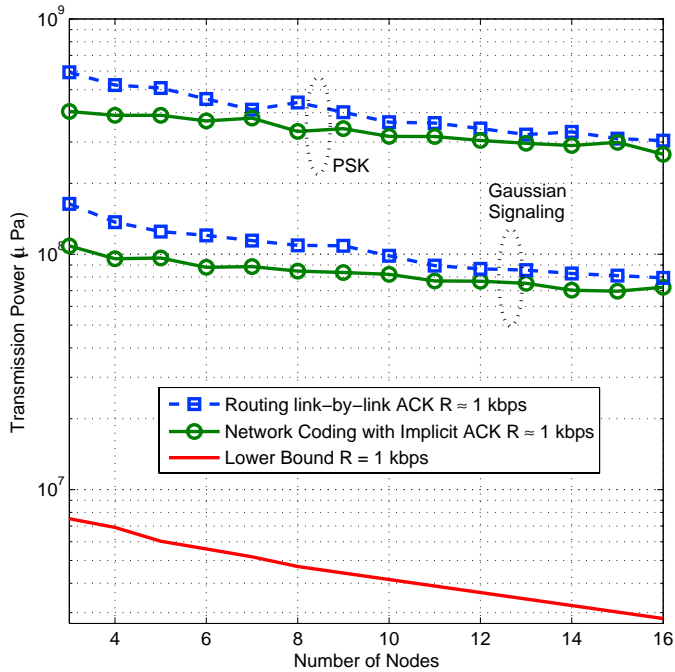}
\caption{Average transmission power of networks deployed randomly on a 1x1~km$^2$ square. Network schemes operating at SNR~=~10~dB and a lower bound for transmission power (scheme 1) is presented.  Model used considers $k=1.5$, $s = 0.5$ and $w = 0$~m/s.}
\label{RoutingNetCodSNR10dB2D.tag}
\end{figure}

	Let us study some numerical results that correspond to a network in which nodes are deployed randomly in a two dimensional space. Unicast connections of rate $R$ are established, i.e. the network has one transmitter, one receiver chosen randomly, and, possibly, several relay nodes. The number of nodes ranges from 3 to 8. The transmission power lower bound, as an average over random deployments, will be compared with transmission power of schemes for routing and network coding. Also, a comparison between the schemes (1) and (2) in terms of interference is presented. Note that the transmission power lower bound is computed assuming that all nodes are within the transmission range of the others, i.e. full connectivity.
	
%	Figure \ref{SameSquare_diff_nodes.tag} shows the lower bound to transmission power applied to three unicast rates $R$ and different number of nodes for the first scheme, considering placing nodes randomly in a 5~x~5~km$^2$ square. The case of two nodes in the graph refers to a simple link. This figure shows that adding an extra node in the network reduces the total average transmission power by 20~\%. However, to reduce transmission power of the link transmission to 40~\% or 60~\% it requires 3 and 6 nodes, respectively, to be added to the network. Thus, it becomes harder to reduce transmission power by 20~\% by adding nodes, that is we have diminishing returns. Of course, there is a tradeoff involved in adding nodes in underwater networks because the cost of each of the nodes may be high. 
	
	Figure \ref{PercentSameSquareDifferentOmega_andSNR.tag} illustrates the effect of introducing a duty cycle $\theta$ (dashed lines) for link transmission under a random deployment in a 5~x~5~km$^2$ square. For $\theta = 1$ in Figure \ref{PercentSameSquareDifferentOmega_andSNR.tag}, which corresponds to scheme 1, note that less than 3~\% of the random deployments cause severe interference (SIR~$<$~3~dB) over at least one link. This corroborates the no-interference assumption used during the analysis to obtain a lower bound for transmission power. Furthermore, this percentage seems to have little dependence on the number of nodes deployed in the network. When a value of $\theta < 1$ is used, Figure \ref{PercentSameSquareDifferentOmega_andSNR.tag} shows that the percentage of deployments with SIR~$<$~3~dB increases for the initial decrements of $\theta$, but decreases as $\theta$ becomes very small (below 1.5\% for $\theta = 0.01$). Although this may seem counter intuitive, introducing a duty cycle causes the link to transmit at a higher data rate when it is active which translates to using more bandwidth and power to achieve that data rate in the underwater channel. Although duty cycle reduces interference by not using the channel continuously, the combined effect with the increased transmission bandwidth and power causes more interference for initial decrements on the value of $\theta$. This is a transient effect, and it has a breaking point for a small value of $\theta$ when the probability of having interference is small. 
	As the value of $\theta$ decreases the transmission power can be shown to increase. This is an expected effect since the cost function is convex and the value of $\theta$ is a constant parameter to all links in this problem.
	
	Figure \ref{PercentSameSquareDifferentOmega_andSNR.tag} shows the results in continuous lines for the third scheme with different SNR requirements. The figure presents the percentage of random deployments that have at least one link suffering from severe interference. For very low SNR, the assumption of no-interference is justified. However, even for SNR~=~-5~dB the percentage of deployments with severe interference for a unicast connection increases dramatically, especially when the number of nodes in the network increases. A similar effect occurs when SNR~=~0~dB. One way to reduce interference while having an SNR requirement is to use a similar approach as is scheme (2), i.e. to have a transmission duty cycle in each of the links. Schemes (4) and (5) show an implementation using an ALOHA MAC protocol, where every link has an associated duty cycle when it has some data to transmit.        
	
	Let us compare the transmission power of scheme (4) and (5) to the lower bound using both PSK and Gaussian signaling in a 1~x~1~km$^2$. Figure \ref{RoutingNetCodSNR10dB2D.tag} shows the average transmission power for different number of nodes in the network, both active and inactive, i.e. before determining the shortest path or solving the subgraph selection problem, with a transmission power computed to obtain a burst SNR~=~10~dB. The average data rate for the different schemes is $R \approx 1 $~kbps. This figure shows optimal signaling (Gaussian signalling) and a PSK modulation, which illustrates that close to 6~dB in the gap between a PSK modulation and the lower bound is due to the choice of the modulation. Notice that the gap between the average transmission power for Gaussian signaling and the lower bound of  $R = 1$~kbps in Figure \ref{RoutingNetCodSNR10dB2D.tag} is about 11~dB for scheme (4) and 13~dB for scheme (5). Also, it shows that this gap is maintained as more nodes are deployed. 
	Some part of the gap is related to the MAC protocol used. Another is related to the 10~dB SNR requirement which is usually used for a practical implementation. 
	
	\begin{figure}[t]
\centering						
\includegraphics[height=3.5in,width=3.5in]{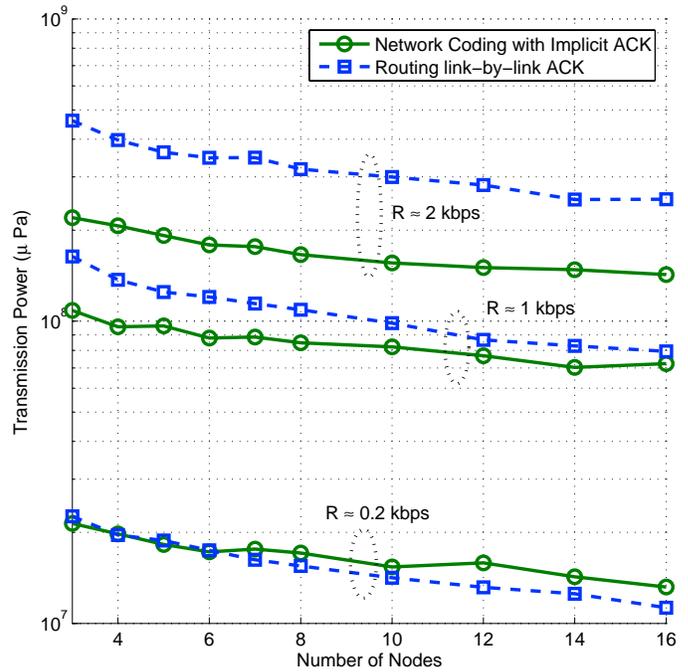}
\caption{Average transmission power of networks deployed randomly on a 1x1~km$^2$ square. Network schemes operating at SNR~=~10~dB using Gaussian signaling. Model used considers $k=1.5$, $s = 0.5$ and $w = 0$~m/s.}
\label{DifferentDataRates.tag}
\end{figure}
\begin{figure}[t]
\centering						
\includegraphics[height=3.5in,width=3.5in]{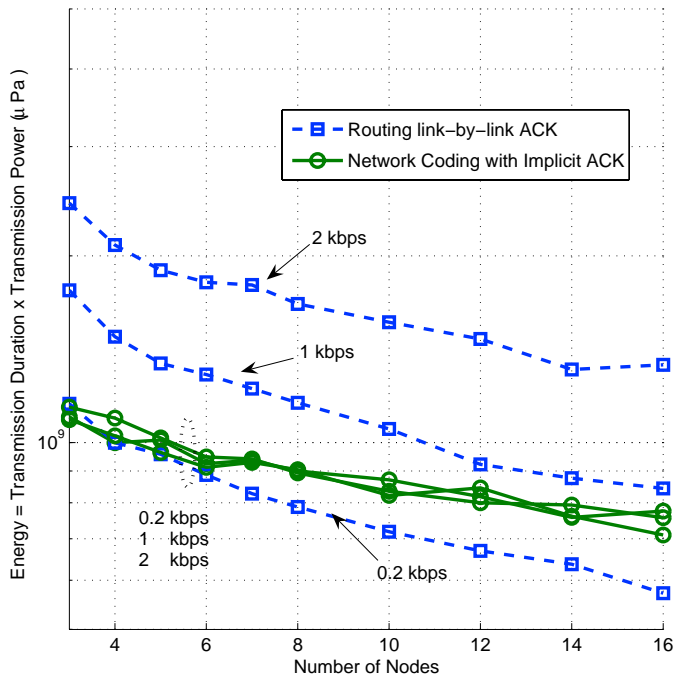}
\caption{Energy for transmission for networks deployed randomly on a 1x1~km$^2$ square. Network schemes operating at SNR~=~10~dB using Gaussian signaling. Model used considers $k=1.5$, $s = 0.5$ and $w = 0$~m/s.}
\label{EnergyDifferentDataRates.tag}
\end{figure}
	
	Figure \ref{DifferentDataRates.tag} compares transmission power for different data rates using Gaussian signaling. The number of transmitted bits was kept constant, while the transmission probability over each link was increased to achieve the desired rate. Note that transmission power increases by 3~dB for scheme (4) while it increases by almost 5~dB for scheme (5) when the data rate is increased from 1~kbps to 2~kbps, i.e. the gap between scheme (4) and (5) increases as data rate increases. This figure shows also that the gap between schemes (4) and (5) is very low when the data rate is 0.2~kbps. Note that an increase in data rate is related to an increase in the collision probability in the ALOHA protocol. For the same setting, Figure \ref{EnergyDifferentDataRates.tag} shows transmission energy for both schemes. The energy required for transmitting at the chosen data rates remains constant in the case of network coding, while it increases for routing when high data rates are attempted. Multiple transmissions of one packet are the main cause of the increased energy consumption for scheme (5), caused both by packet losses and long delays in transmitting an ACK packet given the ALOHA MAC layer. While scheme (4) transmits innovative packets at each transmission, scheme (5) tries to retransmit the same packet if no ACK has been received. Consider the case of a long delay in transmitting an ACK, i.e. the packet was correctly received but the ACK is transmitted a long time after reception, scheme (5) can generate several transmissions of the same data packet. This involves an additional energy consumption. For the same number of transmissions, scheme (4) will transmit several innovative packets, which are useful in decoding the information at the receiver. 
  
	These results illustrate that coding, subgraph selection and the eavesdropping capabilities associated with network coding allow a better performance when the collision probability increases. However, notice that when transmission rates are low the benefits of network coding are less marked. This is explained by the fact that an implicit ACK  might or might not be received by an upstream node. If it is not received, the node will keep transmiting innovative packets. This effect is particularly evident when Gaussian signaling since only nodes in range of transmission will correctly receive a packet. If we use a similar example as in figure \ref{MAC.tag}, node 1 will not receive any implicit ACK from node 2, and it will continue to transmit until informed that all information was received at the sink node. Thus, an explicit ACK procedure should be used if node deployments are likely to produce these situations, especially if only one node is actively generating new data packets.
	
\section{Conclusions}

	A tractable model for the underwater acoustic channel has been established and used for network optimization. A lower bound for transmission power in underwater acoustic networks was obtained based on network coding, and this bound was used to compare the performance of several network coding and routing schemes.

	The closed-form approximate models for the time-invariant acoustic channel where shown to provide a good fit to the actual empirical values by numerical evaluation for different ranges of distance $l$ and capacity $C$, as well as noise profiles corresponding to different shipping activity factor and wind speed. The parameters obtained for these approximate models can be used in the case that a different spreading factor is needed, since the band-edge frequency $\hat{f}_{end}(l,C)$ and the bandwidth $\hat{B}(l,C)$ were shown to be invariant to the spreading factor $k$, while the power scales as $P(l,C,k') = ( l/l_{ref})^{k' - k} P(l,C,k)$, where $l$ is in km. Also, the approximate model of $P(l,C)$ was shown to be almost independent of the shipping activity factor $s$ while having a marked dependency on the wind speed $w$ for $l<10$~km and $C<2$~kbps. This dependence on $w$ is quite smooth and can be approximated by a simple model, resulting in a complete model for the $P(l,C)$ for a range of values $(l,C)$ that is of interest to a typical underwater communication system. 

	We show that the complete model is convex for all $C>0$ and distances $l>0$. Since the complete model is complicated to solve a network optimization problem, we present conditions on the minimum distance between nodes that ensure convexity of the approximate model. Convexity of this model allows us to use it in more complex scenarios, for example, in the framework of layering as optimization decomposition \cite{chiang06} \cite{chiang07}.
	
	This work shows that the no-interference assumption used in the computation of the lower bound to transmission power in the underwater scenario is justified for low multicast rates, and randomly placed nodes with distances under 10 km between each other. We present numerical results that confirm the validity of this assumption, showing that less than 3\% of the links suffer severe interference. We show that solving the optimization problem when the links have a fixed duty cycle for transmissions can reduce interference if the duty cycle is low enough. However, high duty cycles can effectively increase interference because of the high dependence of the bandwidth to the transmitted data rate in an underwater channel. 
	
	The network coding based lower bound was used to determine the gap of different medium access protocols and network schemes for some multicast rate in underwater networks. This comparison is carried out for several routing and network coding schemes. We present numerical results in unicast scenarios by deploying nodes in a two dimensional environment.  

	Network coding with implicit acknowledgements introduced in \cite{lucani07} has better performance than routing with link-by-link ACK in terms of transmission power, especially when the probability of collision is increased. The gap between routing with link-by-link ACK and network coding with implicit ACK in terms of transmission power was shown to increase as the transmission probability increased, which is closely related to the probability of packet collision. However, study of a network coding scheme with an explicit ACK should be considered to improve performance. Also, network coding with implicit acknowledgement compared to common rateless network coding schemes allows to save resources, e.g. memory required in the nodes, and rate adaptation following a similar analysis as in \cite{fragouli07}. Also, there has been previous work on network coding based Ad-Hoc protocols, such as CODECAST \cite{gerla06}, which could be extended to use implicit acknowledgements, and adapted to the underwater acoustic channel.   

	Future research should consider using error exponents \cite{gallager} to compute the error probability when Gaussian signaling is used. This provides a more accurate estimate of transmission power when encoding is performed over a finite number of bits with a data rate approaching capacity. Finally, future research should address the issue of scalability, i.e. study if the results are valid when the number of nodes in the network is increased but maintaining the same node density.

% or

% Use this command to get the appendices' numbers in "A", "B" instead of the
% default capitalized Roman numerals ("I", "II", etc.).
% However, the capital letter form may result in awkward subsection numbers
% (such as "A-A"). Capitalized Roman numerals are the default.
%\useRomanappendicesfalse
%

\appendices
\section{Proof of Lemma 1}
	We consider a set $\Xi $ of bands, each band $i \in \Xi $ having a $f_{end}^i(l,C)$ and $f_{ini}^i(l,C)$ associated to it. Then, $P(l,C) =  \sum _{i} K(l,C)(f_{end}^i(l,C) - f_{ini}^i(l,C))  - \sum _{i} \int _{f_{ini}^i(l,C)}^{f_{end}^i(l,C)}A(l,f)N(f)  df$ and $C = \sum _{i} \int _{f_{ini}^i(l,C)}^{f_{end}^i(l,C)} log_{2} \left (  \scriptstyle \frac{K(l,C)}{A(l,f)N(f)} \displaystyle  \right ) df $. 	Using the Leibniz Integral rule, the fact that $K(l,C) = A(l,f_{end}(l,C))N(f_{end}(l,C))$ and $K(l,C) = A(l,f_{ini}(l,C))N(f_{ini}(l,C))$, that $A(l,f)N(f)$ is independent of $C$, and that the derivative of the sum is the sum of the derivatives, $\scriptstyle \frac{\partial P(l,C)}{\partial C} \displaystyle = \scriptstyle \frac {\partial K(l,C)}{\partial C} \displaystyle \sum_{i} (f_{end}^i(l,C) - f_{ini}^i(l,C))$. Taking the second derivative:
\begin{eqnarray}
& \scriptstyle \frac{\partial ^2 P(l,C)}{\partial C^2} \displaystyle =  \sum_{i}  \scriptstyle \frac {\partial ^2 K(l,C)}{\partial C^2} \displaystyle (f_{end}^i(l,C) - f_{ini}^i(l,C)) \displaystyle \\
&+ \sum_{i} \scriptstyle \frac {\partial K(l,C)}{\partial C} \displaystyle  \left(\scriptstyle \frac{\partial f_{end}^i(l,C)}{\partial C} - \frac{\partial f_{ini}^i(l,C)}{\partial C} \displaystyle \right ) 
\end{eqnarray}
	Taking the derivative of $C$ with respect to itself, using Leibniz Integration Rule and $K(l,C) = A(l,f_{end}(l,C))N(f_{end}(l,C))$ and $K(l,C) = A(l,f_{ini}(l,C))N(f_{ini}(l,C))$, then:
\begin{eqnarray}
1 =  \scriptstyle \frac{1}{ln(2)K(l,C)}  \frac {\partial K(l,C)}{\partial C} \displaystyle \sum_{i} \left ((f_{end}^i(l,C) - f_{ini}^i(l,C)) \right )
\end{eqnarray} 
	Since $K(l,C)>0$ for any $l>0$ and $C>0$ by the physics of the channel and $f_{end}^i(l,C) - f_{ini}^i(l,C)>0,\forall C>0, l>0$ , and the $i$ bands are non/overlapping and $ln(2)>0$ this implies that $\frac {\partial K(l,C)}{\partial C}>0$. Then
$\frac{\partial P(l,C)}{\partial C} = ln(2)K(l,C) > 0, \forall l>0, C>0$. Taking a second derivative to the $C$ expression with respect to itself:
\begin{eqnarray}
& \left ( \scriptstyle \frac{\partial K(l,C)}{\partial C} \displaystyle  \right) ^2 \sum_{i}(f_{end}^i(l,C) - f_{ini}^i(l,C))=\\&  \scriptstyle \frac {\partial ^2 K(l,C)}{\partial C^2} \displaystyle \sum_{i}(f_{end}^i(l,C) - f_{ini}^i(l,C))  \\& + \scriptstyle \frac {\partial K(l,C)}{\partial C} \displaystyle \sum_{i} \left( \scriptstyle \frac{\partial f_{end}^i(l,C)}{\partial C} - \frac{\partial f_{ini}^i(l,C)}{\partial C} \displaystyle \right )
\end{eqnarray} 
	Thus, $\scriptstyle \frac{\partial ^2 P(l,C)}{\partial C^2}  \displaystyle = \left (  \scriptstyle \frac{\partial K(l,C)}{\partial C} \displaystyle \right) ^2 \sum_{i}(f_{end}^i(l,C) - f_{ini}^i(l,C))$ where $(f_{end}^i(l,C) - f_{ini}^i(l,C)) > 0, \forall C>0$, finite and non-overlapping and $\frac{\partial K(l,C)}{\partial C} > 0$. Thus,$\frac{\partial ^2 P(l,C)}{\partial C^2}  >0$
\section{Proof of Lemma 2}
Since $A(l,f) = (l/l_{ref})^k a(f)^l$ and $\frac{\partial A(l,f)}{\partial l} = (k/l_{ref})(l/l_{ref})^{k-1}a(f)^l + (l/l_{ref})^k ln(a(f))a(f)^l >0$ since $a(f)\geq 1$ and $l_{ref}>0$. Then, $A(l,f) >A(l',f), l>l'$. 
Also, $\frac{K(l,C)}{A(l,f)N(f)} \geq 1, \forall f \in B(l,C)$ which implies $log_{2}(\frac{K(l,C)}{A(l,f)N(f)}) \geq 0, \forall f \in B(l,C)$. Let us compute the capacity of a link of distance $l'$ when we use the optimum band and spectral density for a link of distance $l$ and capacity $C$, i.e. $B(l,C)$ and $S(l,C,f) = K(l,C) - A(l,f)N(f), f \in B(l,C)$, respectively. Then, 
\begin{eqnarray}
&C(l',B(l,C)) = \int _{B(l,C)} log_{2} \left (   1 + \scriptstyle \frac{K(l,C) - A(l,f)N(f)}{A(l',f)N(f)} \displaystyle  \right ) df\\& > \int _{B(l,C)} log_{2} \left (   1 + \scriptstyle \frac{K(l,C) - A(l,f)N(f)}{A(l,f)N(f)} \displaystyle  \right ) df = C
\end{eqnarray}

%\section*{Acknowledgment}
% optional entry into table of contents (if used)
%\addcontentsline{toc}{section}{Acknowledgment}



% http://www.ctan.org/tex-archive/macros/latex/contrib/supported/IEEEtran/bibtex
%\bibliographystyle{IEEEtran.bst}

\singlespacing

%\begin{biography}[{\includegraphics[width=1in,height=1.25in,clip,keepaspectratio]{mshell}}]{Michael Shell}
% where an .eps filename suffix will be assumed under latex, and a .pdf suffix
% will be assumed for pdflatex; or if you just want to reserve a space for
% a photo:

\begin{biography}[{\includegraphics[width=1in,height=1.25in,clip,keepaspectratio]{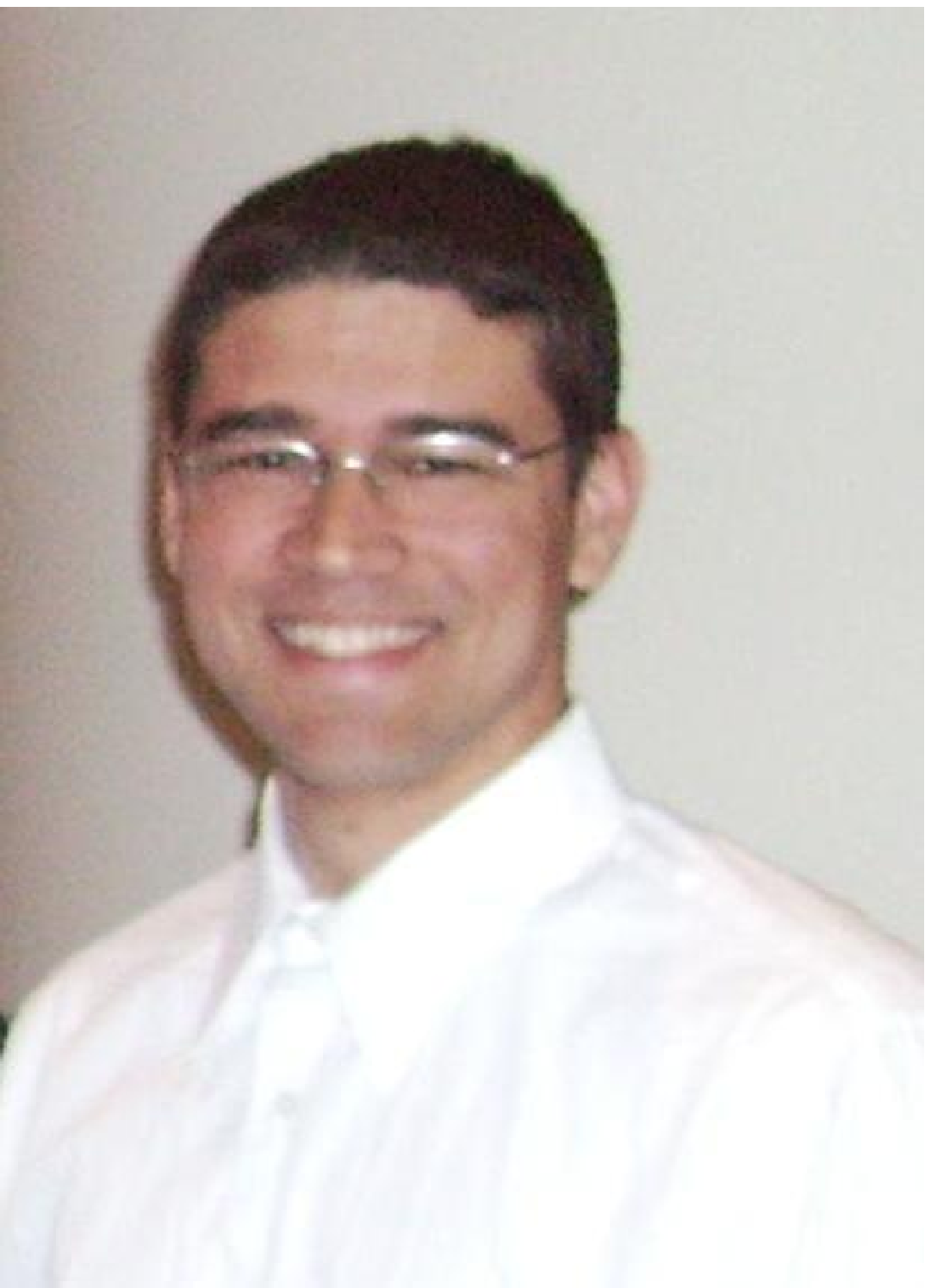}}]{Daniel Lucani}
received the B.S. degree in Electronics Engineering (\textit{summa cum laude}) and the M.S. in Electronics Engineering (with honors) from Universidad Sim\'on Bol\'ivar, Venezuela, in 2005 and 2006, respectively. He is currently a Ph.D. candidate at Massachusetts
Institute of Technology (MIT). His research interests include digital communications, wireless communications and networks, network coding, and their applications to mobile radio and underwater acoustic communications.
\end{biography}
\vfill
\begin{biography}[{\includegraphics[width=1in,height=1.25in,clip,keepaspectratio]{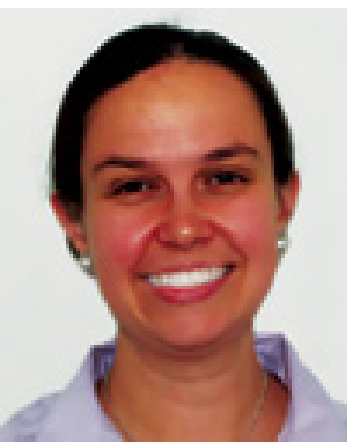}}]{Muriel M\'edard}
is a Professor in the Electrical Engineering and Computer Science at MIT.
She was previously an Assistant Professor in the Electrical and Computer Engineering
Department and a member of the Coordinated Science Laboratory at the University of
Illinois Urbana-Champaign. From 1995 to 1998, she was a Staff Member at MIT Lincoln
Laboratory in the Optical Communications and the Advanced Networking Groups. Professor
M\'edard received B.S. degrees in EECS and in Mathematics in 1989, a B.S. degree in
Humanities in 1990, a M.S. degree in EE 1991, and a Sc D. degree in EE in 1995, all from
the Massachusetts Institute of Technology (MIT), Cambridge. She is an associate editor
for the IEEE Journal of Lightwave Technology. She has served as an Associate Editor for
the Optical Communications and Networking Series of the IEEE Journal on Selected Areas in
Communications, as an Associate Editor in Communications for the IEEE Transactions on
Information Theory and as an Associate Editor for the OSA Journal of Optical Networking. 
She has served as a Guest Editor for the IEEE Journal of Lightwave Technology, the Joint
special issue of the IEEE Transactions on Information Theory and the IEEE/ACM
Transactions on Networking on Networking and Information Theory  and the IEEE
Transactions on Information Forensic and Security: Special Issue on Statistical Methods
for Network Security and Forensics. She is a member of the Board of Governors of the IEEE
Information Theory Society.
Professor M\'edard's research interests are in the areas of network coding and reliable
communications, particularly for optical and wireless networks. She was awarded the
IEEE Leon K. Kirchmayer Prize Paper Award 2002 for her paper,``The Effect Upon
Channel Capacity in Wireless Communications of Perfect and Imperfect Knowledge of the
Channel," IEEE Transactions on Information Theory, Volume 46 Issue 3, May 2000, Pages:
935-946. She was co- awarded the Best Paper Award for G. Weichenberg, V. Chan, M.
M\'edard,``Reliable Architectures for
Networks Under Stress", Fourth International Workshop on the Design of Reliable
Communication Networks (DRCN 2003), October 2003, Banff, Alberta, Canada. She received a
NSF Career Award in 2001 and was co-winner 2004 Harold E. Edgerton Faculty
Achievement Award, established in 1982 to honor junior faculty members ``for distinction
in research, teaching and service to the MIT community." She was named a 2007 Gilbreth
Lecturer by the National Academy of Engineering. Professor M\'edard is a House Master at
Next House and a Fellow of IEEE.
\end{biography}
\begin{biography}[{\includegraphics[width=1in,height=1.25in,clip,keepaspectratio]{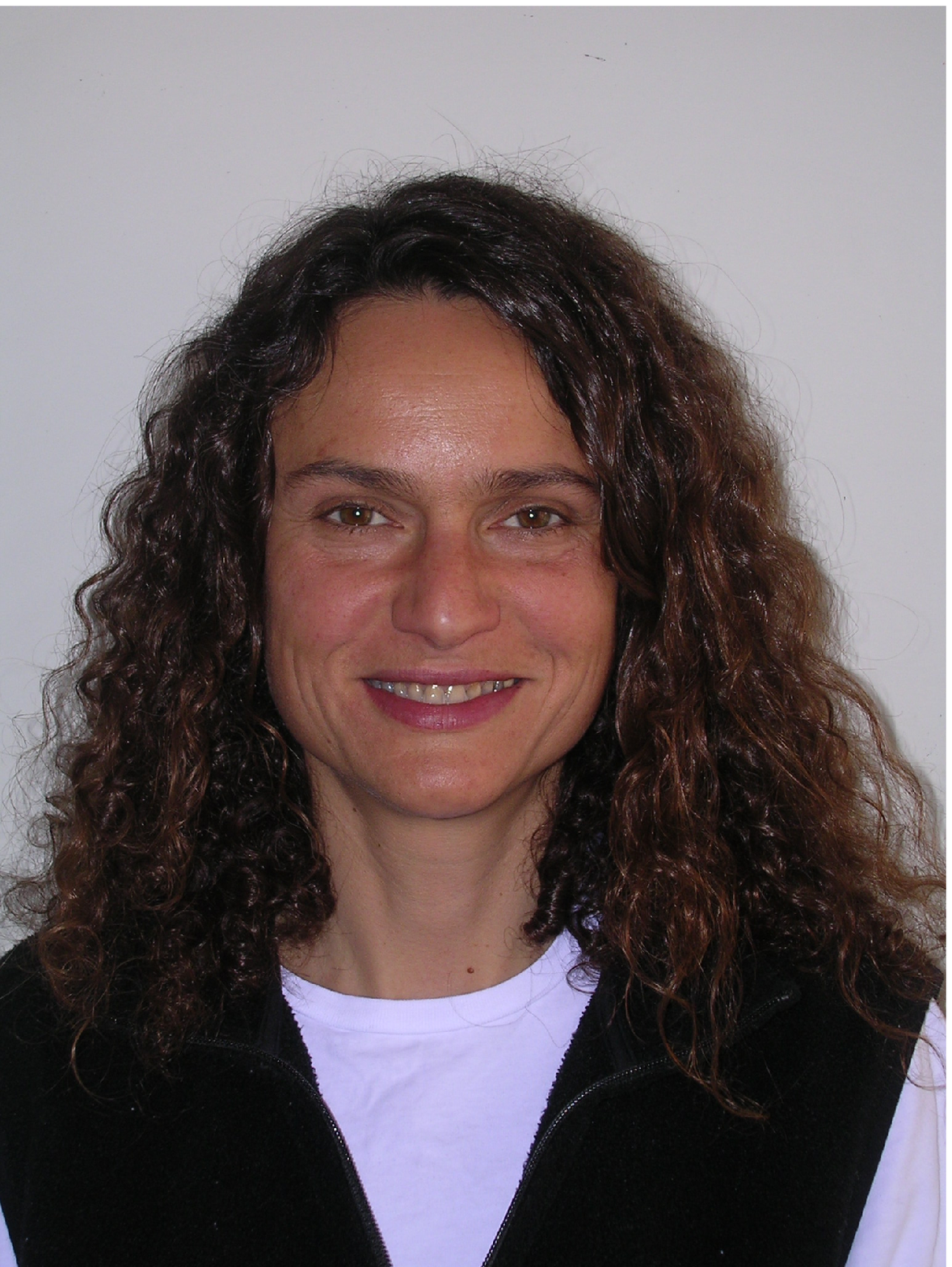}}]{Milica Stojanovic}
graduated from the University of Belgrade, Serbia, in 1988, and
received the
M.S. and Ph.D. degrees in electrical engineering from Northeastern
University, Boston, MA, in 1991 and 1993. After  a number of years with the Massachusetts
Institute of Technology, where she was a Principal Scientist, in 2008 she joined  the
faculty of Electrical and Computer Engineering Department at Northeastern University. She
is also a Guest
Investigator at the Woods Hole Oceanographic Institution, and a Visiting Scientist at MIT.
Her research interests include digital communications theory, statistical
signal processing and wireless networks, and their applications to mobile radio and
underwater
acoustic communication systems.
\end{biography}
\vfill


%\vfill

%\enlargethispage{-5.0in}
\IEEEtriggercmd{\enlargethispage{-5.0in}}

%\begin{figure}[p]
%\centering						
%\includegraphics[height=3.5in,width=3.5in]{lnLlowC.pdf}
%\caption{ Region of $(l,z)$ where the approximate model for $k=1.5$, $s = 0.5$ and $w = 0$~m/s is convex with respect to $z$. The model is convex $\forall (l,z)$ over the curve. Note that for $l>13 $~m the model is convex $\forall z$ of interest.}
%\label{lnLlowC.tag}
%\end{figure}    

%\begin{figure}[p]
%\centering						
%\includegraphics[height=3.5in,width=3.5in]{SameSquare_diff_nodes.pdf}
%\caption{Minimum transmission power of network in a fixed square of 5x5~km$^2$ vs number of nodes deployed in that area, with different data rates. Model used considers $k=1.5$, $s = 0.5$ and $w = 0$~m/s. }
%\label{SameSquare_diff_nodes.tag}
%\end{figure}

\end{document}